# The academic motherload: Models of parenting engagement and the effect on academic productivity and performance

**Gemma E. Derrick[1*], Pei-Ying Chen[2], Thed van Leeuwen[3], Vincent Larivière[4,5], Cassidy R. Sugimoto[6]**

[1]Department of Educational Research, Lancaster University, United Kingdom
[2]School of Informatics, Computing, and Engineering, Indiana University Bloomington, USA
[3]Centre for Science and Technology Studies, Leiden University, The Netherlands
[4]École de bibliothéconomie et des sciences de l'Information, Université de Montréal, Montréal, Canada
[5]Observatoire des sciences et des technologies, Université du Québec à Montréal, Canada
[6]School of Public Policy, Georgia Institute of Technology, USA

*Gemma Derrick
**Email:** g.derrick@lancaster.ac.uk

**Abstract**

Gender differences in research productivity are well documented, and have been mostly explained by access parental leave and child-related responsibilities. Those explanations are based on the assumption that women take on the majority of childcare responsibilities, and take the same level of leave at the birth of a child. Changing social dynamics around parenting has seen fathers increasingly take an active role in parenting. This demands a more nuanced approach to understanding how parenting affects both men and women. Using a global survey of 11,226 academic parents, this study investigates the effect of parental engagement (Lead, Dual (shared), and Satellite parenting), and partner type, on measures of research productivity and impact for men and for women. It also analyzes the effect of different levels of parental leave on academic productivity. Results show that the parenting penalty for men and women is a function of the level of engagement in parenting activities. Men who serve in lead roles suffer similar penalties, but women are more likely to serve in lead parenting roles and to be more engaged across time and tasks. Taking a period of parental leave is associated with higher levels of productivity, however the productivity advantage is lost for the US-sample at 6 months, and at 12-months for the non-US sample. These results suggest that parental engagement is a more powerful variable to explain gender differences in academic productivity than the mere existence of children, and that policies should that factor into account.

# Introduction

The COVID-19 pandemic placed the productivity cost of parenting into sharp relief. Studies provide evidence of a decrease in women first-authorship on preprints (1, 2), papers (1), and funding applications (17), and lower participation in academic citizenship activities (3) than men. Consistent with previous research on gender disparities in science (1,2), reasons for this inequity reference women's engagement in household responsibilities (23, 3) and an increase of caregiving activities during COVID-19 related lockdowns, both for academics (4) and non-academics (5, 6). The argument is simple: childcare and homeschooling for extended periods of time were overwhelmingly fulfilled by women (7, 8) thereby decreasing their engagement in professional responsibilities.

These inequities, however, predate the pandemic. Decades of work on gender in science have shown that, on average, women publish fewer articles (9–11) and receive fewer citations for their work (12), even when publishing in journals with higher Impact Factors (13, 14). Domestic responsibilities associated with parenting have been heralded as an explanation for these differences (15); yet, studies on the relationship between parenting and scientific productivity are mixed. Studies have found that women with children face a productivity penalty compared to men with children and women without children (16–19). Other studies have suggested no association between production and family obligations (20) or an increase in productivity immediately after birth (17, 21). The latter may be an artifact of publication lags and the pressures of an academic environment: demonstrating that women accelerate their productivity directly *before* birth, which manifests after.

Many of the explanations belie the complexities of modern parenting (21). Women have entered the workforce at larger rates, changing the norm to a dual-labor household (24, 25). In addition, the concept of fatherhood has also changed with more fathers taking an increasingly central role in child-rearing (32) even during the pandemic (33, 34). Studies also suggest attrition from STEM after the birth of a child for fathers (31). These new household dynamics has increased the use of several adaptation behaviors, including labor shifts between parents (25) and the utilization of extended family members for primary childcare (22–24). In addition, to characterize the balance of labor, it is essential to examine not only the workweek, but also the weekends (25), particularly for researchers, who tend to have non-standard work schedule. Furthermore, studies often assume that having a child is equated with homogeneous engagement with children, across parents. This assumption fails to take into account unequal parenting responsibilities, and the inclusion of multiple caregivers in a household. More nuanced research designs are necessary to understand how parental engagement influences research productivity and visibility beyond the current binary categorization of parenthood.

To understand the complex relationship between parenting and academic work, we employ a mixed methods approach to consider how different models of parenting engagement and household arrangements impact academic productivity and impact for men and women. We address the following research questions:

1. Is there a gendered difference in parenting engagement for researchers?
2. Is there a relationship between parental engagement and research productivity? If so, does this differ by gender and household composition?
3. What is the relationship between degrees of parental leave and productivity by gender? If so, is this mediated by gender and household composition?
4. Is there a gender difference between parenting engagement and scientific impact? If so, is this mediate by gender and household composition?

**Methods**

***Population and sample.*** Population for this survey was defined as any first or last authors of 2007-2016 papers indexed in Clarivate Analytics' Web of Science (WOS), for which an email address could be retrieved (N=2,640,872). A random sample of 1.5 million researchers was generated from this population, to which the survey was sent in 15 batches of 100,000 email invitations through Qualtrics between February 15 and March 8 of 2018. Survey questions asked respondents about the number of children, the years of birth, if and how much maternity/paternity leave was taken following the birth of each child, as well as their engagement in activities that were related to providing care/decision-making for children. In addition, demographic information on children and partners of respondents were included, as well as contribution to childcare and academic careers. In total, 17,519 researchers responded, of which 14,910 were considered eligible who finished survey and reported having children. Further data processing was performed to exclude incomplete and anonymous responses, reducing the sample size to 11,226, or 75.3% of eligible responses.

This subset of 11,226 respondents was then matched with their corresponding publication records in the WOS over the 1980–2017 period. Such publication records were automatically disambiguated using the algorithm developed by Caron & van Eck (35), which uses heuristics such as researchers' field of study, institution of affiliation, collaborators and cited references to automatically reconstruct researchers' publication records and distinguish papers written by two or more authors that share the same name. This algorithm has been shown to have high precision and recall (36), and to produce the best results among the existing disambiguation algorithms (37). There were, however, respondents to which we were unable to match a publication record, and researchers for which no citation information was found were also excluded for the analysis (N=781 researchers). On the whole, the final sample analyzed contained 10,445 researchers, which represents 70.1% of all researchers who have finished the survey. This accounts for 0.70% of the set of sampled researches, and for 0.40% of the entire population.

***Quantitative Analysis.*** The analytic set for multiple linear regression models includes 10,013 respondents after excluding 432 with zero MNCS. The outcome variables are the number of publications indexed in WOS between 1980 and 2017 in the productivity model. For the impact model, both the number of total citations (TCS) and mean normalized citation scores (MNCS) are used as proxies for research impact. Natural log transformation was applied to normalize the indicators. We identified three parenting types based on respondents' caregiving situation (either current or when their children were dependents). Lead parents are the primary caregivers to their children, dual parents share equal parenting roles with either their partner or non-parental caregivers (e.g., grandparents, nannies), and satellite parents whose partner or non-parental others are primary caregivers for children. The lengths of parental leaves are divided into six categories: no leave, less than one month, at least one but less than three months, three to six months, more than six but less than 12 months, and 12 months or more.

To test whether the relationship between parental engagement and research performance differs by gender and partnership status, we include a three-way interaction term between parenting type, partnership status, and gender. Partnership status is determined by the current occupation of respondents' partner to be in an academic or non-academic sector. Academic partner is strictly referred to respondents whose partner is employed in academic sectors for research and/or teaching. Partners employed in government, private, or other sectors, as well as retired and not employed, are all designated as non-academic partner. To make our analysis more inclusive, single parents are assumed to be no partner and retained in the sample.

Considering the strong presence of the US respondents in our sample and the fact that US is the only advanced country without a nation-wide parental leave policy, we split the sample into the US-only and non-US subsets when modeling the effect of parental leave lengths on productivity. In addition to the above-mentioned three-way interaction between parenting type, partnership status, and gender, we also include an interaction term between gender and parental leave lengths to examine whether the relationship between parental leave lengths and productivity differs by gender. We control for the number of children as well as respondents' academic age, highest degree earned, employment sector, and primary discipline. Academic age is calculated based on the first and last years of publication plus one. Given that the effect of age is often multiplicative, we include academic age in its polynomial form and center the variable at the mean to render the intercept more meaningful and interpretable. Both the highest degree and employment sector are binary variables, indicating whether respondents hold a doctoral degree and work in an academic sector for research and/or teaching. Number of children is a nominal variable with four possible values: 1, 2, 3, as well as 4 and more; so is the main discipline, which is regrouped from the original 14 disciplines into four: Arts and Humanities, Health Sciences, Natural Sciences, and Social Sciences.

***Qualitative analysis.*** A free text section was included at the end of the survey that encouraged participants to *"Please feel free to add any additional comments you have regarding childcare and scientific labor, drawing upon your own experiences"*. In total, 5,976 participants completed this section. To analyze this, a random sample of 1500 was selected and inductively coded into themes using a grounded theory-informed approach (38, 39). All codes were then collapsed into 59 overarching axial codes capable of coding a large number of responses. An additional 'Other' category was permitted to retain data variability and richness. The remainder of responses (n=4,476) were then manually coded into these themes.

**Results**

***Parenting labor by gender.*** Among our respondents, women are more likely than men to serve as the primary caregiver for their children (30.6% vs. 3.9%) (Figure 1A). Inversely, more than a third of men (38.9%) indicated that they served in a secondary role in parenting (i.e., satellite), compared to only 17.4% of women. This establishes an important baseline for the study of scientific parenting: that is, women scholars are disproportionately likely to be taking a lead role in caregiving. The most common model, however, is one of shared parenting: the majority of men (57.1%) and women (52.0%) indicated that parenting roles were shared equally with their partner (i.e., dual).

These self-reported roles were investigated to understand how they manifest themselves in terms of both time and task engagement—that is, the times during which parents were engaging in parenting and the types of tasks with which they were disproportionately associated. Lead parents of both genders indicated high percentages of engagement across time compared to other roles, suggesting that the 3.9% of men serving in this role are strongly engaged in parenting. However, men in dual and satellite parenting roles were much less likely than women in the same roles to report primary caregiving across times. This demonstrates that there is a higher burden of labor for women to classify themselves as dual and satellite parents than men (Figure 1B). Men reinforced these disproportionate labor expectations, even in shared parenting relationships:

> *Although I try to be active in child care and share responsibilities equally, my wife still takes care of more child care tasks than I do.* (M, Dual, US)

The time results were confirmed by an analysis of tasks. In nearly every category—particularly for dual and satellite parents—women were more likely to be the parent conducting caregiving tasks (Figure 1C, Table S1). There were fewer differences in how men and women operationalized lead parenting, with lead parents of both genders significantly engaged in childcare tasks. The only task where lead fathers demonstrated significant differences were in dropping off children at school/nursery (79.3% vs. 66.4%, $\chi^2$=12.12, $p$=0.001) and coaching sports (40.8% vs. 17.7%, $\chi^2$=53.66, $p$<0.001). Much stronger gender differences were observed in dual parenting. Men in dual roles were more likely to drop off children at school/nursery (40.0% vs. 29.3%, $\chi^2$=70.98, $p$<0.001), coach sports (30.5% vs. 5.6%, $\chi^2$=625.04, $p$<0.001), and do school/nursey pick-up (28.7% vs. 27.0%, $\chi^2$=2, $p$=0.168), though the latter was not significant. Women in dual roles were significantly more likely to be primarily responsible for all other caregiving duties. The same was true for satellite roles. The only task with which men were significantly more likely to be associated was coaching sports (25.8% vs. 5.1%, $\chi^2$=188.44, $p$<0.001).

The time and task analyses reinforce each other: when they self-identified as dual or satellite parents, women are disproportionately engaged in parenting activities. Furthermore, there is little difference between women's labor performance in dual and satellite roles (Figure S1). These asymmetries between labor and credit show that, even in the perception of equality, women are carrying a higher burden of labor.

Qualitative responses were illuminating in this regard. A woman from Tunisia noted that the survey made her aware that she was the main caregiver. Other respondents supported this, but questioned the exhaustivity of the task list:

> *This survey was instructive as I didn't realize how many tasks I take on compared to my partner. I thought it was more equal. Maybe there were tasks not listed. For example, he handles administrative tasks like keeping track of bank accounts, keeping the printer ink filled, and other tasks like basic cleaning within specific tasks (e.g., loading the dishwasher) and yard work.* (W, Dual, US)

The incompleteness of the list was also questioned by other respondents, but in the opposite direction, calling into focus the unequal cognitive and emotional labor performed by women:

> *There is a huge cost to something that you didn't ask about: "running the house" - it's not just child care. It's scheduling someone to come clean and/or cleaning, getting groceries, scheduling sitters, arranging for travel, paying bills, sorting the mail, getting the kids new clothes etc. etc. Often doctor's offices etc. have to be called during business hours and that takes away work time. It's much much more emotional work because I "keep track" of everything. We are traveling for work and need vaccines -- I am the one coordinating them. My husband helps, and definitely the physical aspects of taking care of the kids are 50/50 at least, but there's all this other stuff.* (W, Dual, US)

The cognitive and emotional burdens of domestic labor disproportionately born by women have been well-recognized in previous studies (26) (e.g., Ward and Wolf-Wendell, 2004) and were manifest in the exogenous shock of the pandemic (2, 4). Therefore, the inequities observed in the itemization of task and time may only represent a conservative estimate of the actual difference in parenting engagement. However, our work reinforces that self-report data of shared parenting discounts women's engagement in parenting and overestimates men's.

***Work arrangements and adaptation in parenting.*** One limitation of the survey is that it captured individual nodes in dyadic relationships, rather than paired couples. One might expect, for example, different labor roles based on the occupation of the partner. To control for this, we identified the sector of employment of the partner, with a particular focus on situations where both the respondent and their partner were employed in academia. Academic couples arguably experience the same productivity pressures and job responsibilities as each other, creating a natural control for labor expectations. Overall, academic women as dual parents are strongly affected by having an academic partner: in terms of the task analysis, women with non-academic partners are primarily responsible for a larger number of tasks than their counterparts with academic partners—especially regarding transportation and evening care (Figure 2, Table S2). In contrast, academic men are much less affected by their academic partnership status except for dual parents—those with an academic partner are more likely to do household shopping (23.9% vs. 18.8%, $\chi^2$=7.9, $p$=0.038) but less likely to do the school run (drop-off: 35.2% vs. 44.8%, $\chi^2$=19.37, $p$=0.002; pick-up: 22.7% vs. 30.0%, $\chi^2$=13.91, $p$=0.004), put children to bed (14.7% vs. 19.3%, $\chi^2$=7.61, $p$=0.042), and read bedtime stories (26.6% vs. 21.8%, $\chi^2$=6.34, $p$=0.043). For academic men as either lead or satellite parents, we found no statistically significant differences between those with and without an academic partner in parenting engagement (Figure 2, Table S3).

The benefits of a partner who is understanding of and aware of the labor burdens of academe was evident in the qualitative responses. For example, parents commented on the perceived flexibility of research careers and these advantages were enhanced when both parents were academics:

> *"...My wife is also an academic which actually helped in sharing duties in a much more understanding way". (M, Dual, Singapore)*

However, whereas an academic career was seen as more flexible and therefore amenable to parenting, there was an assumption that flexibility also implied availability. This was most evident in respondents who were the sole academic in their household and the delicate balance between flexibility and availability was experienced by both women (*"Inevitably, we both feel that if a sacrifice must be made, it is my schedule" (W, Dual, US)*) and men;

> *"It is hard to balance academic work and home life - as in many cases your partner does not understand that reading and working on your computer is your job. Thus, you find that you have various tasks (family, children, house, errands) thrown to you by your spouse who works a "regular" job because you are "not busy". (M, Dual, US)*

Women, however, are particularly affected by partner occupation, with significant differences in equity between those with and without academic partners. This suggests that spousal hiring programs may have stronger implications for women faculty in caregiving roles as the equitability of parenting tasks is higher with academic partners.

***Effect of parental engagement on research productivity.*** Although parenting engagement and partner types account for only a small fraction of variations in productivity (after controlling for academic age, number of children, and discipline), certain patterns are revealing. As illustrated in Figure 3, both men and women suffer a productivity loss when they are single or lead parents. Using dual mothers with a non-academic partner as our reference group, dual fathers ($\beta=0.05$, *p*=0.029) and satellite mothers ($\beta=0.09$, *p*=0.004) are 5.6% and 8.9% more productive, respectively; single mothers are 15.3% less productive ($\beta=-0.17$, *p*=0.048) (Table S4). The differential effects of parenting engagement involving partnership status for men and women is confirmed in the two-academics subsample, where lead parents are, on average, 11.1% less productive than dual parents ($\beta=-0.12$, *p*=0.012), and the magnitude is roughly the same as the additional effects of being a lead father and being a lead mother with an academic partner in the full sample. This suggests that parenting penalties are felt by both men and women. As one respondent noted:

> *My scientific productivity has declined since having kids- something had to give. I still work hard to be a good teacher and leader in my department while doing as much parenting as possible so my wife can pursue her career goals as well. The cost of the parenting has largely been a reduction in papers written.* (M, Lead, US)

Despite the positive but not significant effect on productivity of having an academic partner for dual mothers ($\beta=0.05$, *p*=0.082), we saw an additional 10% decrease in productivity

(β=−0.11, *p*=0.038) for lead mothers with an academic partner. This is tied with the reference point of dual mothers, where we see that dual mothers operate at similar levels of engagement as lead mothers. Perhaps counterintuitively, lead-mothers are as productive as dual mothers when they both have a non-academic partner (β=0.00, *p*=0.862). This is in sharp contrast to lead fathers who suffer an additional 11.6% decrease in productivity (β=−0.12, *p*=0.11), although not significant. This is likely a result of the unequal engagement women demonstrates in these parenting roles. Men seem to be most productive when they are in satellite roles with academic partners. Women, on the other hand, are most productive when they are in a satellite role with a non-academic partner. This may be a result of the perceived flexibility of academic roles intersecting with cultural expectations in parenting. As one woman observed:

> *You have to be prepared to work twice as hard and accept that. (W, Dual, UK)*

***Effects of parental leaves on production.*** Parental leaves are associated with higher production, but have a point of diminishing return that varies by country (Figure 4, Table S5). The production advantage is highest at less than one month of leave for the US sample (estimated increase of 26.9%; β=0.24, *p*=0.002) and decreases for every three months of leave after (26.7% increase for longer than one but less than three months [β=0.24, *p*<0.001], and 17.8% for three to six months [β=0.16, *p*<0.001] among US women). The advantage disappears after six months with the US sample and after twelve months in the non-US sample. These cultural differences may be explained by the normative leave lengths and productivity expectations by country. One woman from the US explained how the casual terminology reinforced expectations of production during the limited leave given to US mothers:

> *The six weeks after giving birth should be termed medical leave for the person who delivered the baby (regardless of if they are parenting the child). That should be treated as such. There are still people who will call it a 'sabbatical'. (W, Dual, US)*

While the effect is relatively modest in the non-US sample and non-significant for taking leaves less than one month (β=−0.07, *p*=0.609), the corresponding increase in number of papers is estimated to be 17.1%, 10.5%, and 10.6% for leaves longer than one but less than three months (β=0.18, *p*<0.003), three to six months (β=0.10, *p*=0.009), and longer than six but less than twelve months (β=0.10, *p*=0.008). None of the interaction terms between gender and parental leave length are statistically significant, suggesting that the effect of parental leave length on production does not differ by gender. However, women are more likely to take leave and to take longer leave, leading to a disproportionate production disadvantage.

Respondents noted that family friendly policies did little to mediate the effects that labor demands on scientific production, including the necessity of managing ongoing growth of measures of academic impact over the lifetime of the career:

> "Family friendly policies are all very well but basically just allow you to take time off work; they don't reduce the amount of work that there is to do or remove deadlines". (W, Lead, UK).

***Effect of parenting engagement on scientific impact.*** Two indicators of scientific impact are used to examine the relationship between parenting academic capital: number of total citations (TCS) and mean normalized citation scores (MNCS). While the first one is largely dependent on researchers' number of papers, the second is scale-independent, and indicates the average impact of their paper compared to other papers published in the same specialty within the same year. Results from the TCS and MNCS models are similar in magnitude and significance, although some notable differences exist (Figure 5, Table S4). The same differential effects of having an academic partner and being a lead parent for men and women on productivity is also present for impact. Having an academic partner for a dual mother increases impact by 11% (MNCS) ($\beta=0.10$, $p=0.001$) and 15.3% (TCS) ($\beta=0.14$, $p=0.004$); whereas having an academic partner for a lead mother brings an additional decrease to impact by 17.1% (MNCS) ($\beta=-0.19$, $p=0.002$) and 19.9% (TCS) ($\beta=-0.22$, $p=0.022$). In other words, the effect of being a lead parent for women is moderated by partner's employment type in both production and impact.

The positive effect of being a satellite mother is only significant in the MNCS model ($\beta=0.08$, $p=0.019$). The notable discrepancies between the two models relate to the moderating effect of gender on different aspects of the relationship between partnership status, parenting type, and impact. A dual father with an academic partner decreases MNCS by 9.6% ($\beta=-0.10$, $p=0.031$), whereas a lead father faces an additional 29.3% drop in TCS ($\beta=-0.35$, $p=0.025$). The former can be translated into the non-effect of academic partners for men and the latter the differential effects of being a lead parent for men and women. More specifically, being a lead parent has a negative effect on impact (TCS) for men regardless of their partner's occupation, but for women only if their partner is an academic. This again is confirmed in both MNCS and TCS models fitted with the two-academics sample, where the lead parent effect for women is close to the interaction effect between lead parent and academic partner in the models with the full sample.

Scientific impact is a function of visibility: work is more likely to be cited when authors are visible in the scientific community through collaboration, travel, and other forms of engagement. Therefore, it stands to reason that parenting demands that reduce visibility will translate into lower citation rates. Respondents often discussed how institutional policies were inadequate in compensating for demands of research careers (e.g., the necessity to travel, overnight stay and long, after-hours work). However, partners who are flexible and supportive were essential for engagement in the scientific community:

> *"Flexible work hours are a blessing, but the travelling required for a successful career (conferences, networking, field work) is a nightmare. I am lucky to have a supportive partner, without whom I would not have been able to pull it off". (W, Dual, Sweden)*

**Discussion and Conclusion**

Parenting engagement is related with decreased research productivity and impact; however, the composition and management of the household plays an important role in mediating this effect. Results show that the parenting penalty for men and women is amplified by their level of engagement in parenting activities. Differential participation in parenting may largely explain observed gender effects. Men who serve in lead roles suffer similar penalties, but women are more likely to serve in lead parenting roles and to be more engaged across time and tasks. In addition, despite respondents indicating that they engage in a dual parenting style, women still engaged in a significantly higher level of daily parenting tasks than men, which may explain the divergence in penalties between men and women in these roles. Women simply bear a higher burden of "reproductive labor" (Data Feminism, p178 (27)). Fathers suffer productivity penalties as their engagement increases; however, these penalties are felt more by women given that they are more likely to serve in lead roles and are more engaged in parenting, even when they report dual or satellite parenting styles.

Our analysis offers a novel lens by examining the cost of parenting engagement, as opposed to previous research that focuses on the binary existence of children as a reason for productivity disparities (16–18, 28). Our work therefore provide insights on some of the unexplained productivity differences observed in earlier research which focused merely on the existence of but not engagement with children (29). Results suggest a zero-sum game between productivity and parenting: the more engagement with parenting, the lower the productivity. Policies should account for this by creating greater permeability between work and home life to allow for parents to move more seamlessly between these professional and parenting responsibilities. For example, academic institutions should provide lactation rooms and on-campus childcare, funding agencies should provide provisions allowing PIs to be accompanied by their children and caregivers, and conferences should be restructured to allow parents to fully engage in all activities.

This disproportionate engagement in parenting is exacerbated by the sector of employment of co-parents. There is greater equity in the distribution of parenting tasks when academic women are in dual partnerships with another academic, arguably because the labor expectations for both are shared. Women in dual partnerships with non-academics do far more labor than their counterparts with academic partners. These inequities manifest in differences in productivity. Men and women scholars have lower productivity when they are single-parents and in lead parenting roles. Sharing parenting roles increases productivity for both, but women's productivity is mediated by the occupation of their spouse: those with academic partners see a greater productivity gain than those without. Of course, non-academic roles are much more heterogeneous than academic roles, limiting our ability to interpret the cause of these differences (which operate in opposite ways for lead and satellite roles). The qualitative analysis of open-ended responses shed light on these tensions: academic work is seen as more flexible, therefore causing increased parenting burdens when academics are in partnerships with non-academics. This suggests that spousal hires are an important tool in achieving equity.

The analysis of parental leaves reinforces the notion of a zero-sum game in academic work for US respondents: the longer the leave, the lower the productivity. These differences, however, were not observed for non-US respondents. This could be due to the heterogeneity of parental leave policies across nations (30). While the leave penalty is gender neutral (in that it applies to all) our study follows earlier analysis in showing that mothers are disproportionately likely to take leave (19, 30) and to take longer leave than fathers. This creates an obvious tension: women place a high value on adequate leave policies when selecting an institution (19); however, their use of these leaves places them at a disadvantage in terms of productivity. From a policy perspective, academic labor and opportunity to gain success during an academic career largely reflects the pipeline model, with adherence to "ideal worker norms" (29). A recent study even referred to 'single ladies' as the ideal-type of academic during COVID-19, due to the absence of responsibilities that would otherwise impede progress (31). Given that pipeline model presents barriers to re-entry for those who deviate from this norm or suspend their progression (32), family friendly policies—such as stopping the clock for tenure and longer maternity leaves—that reinforce time away from research can lead to negative effects on careers. Leave policies, therefore, should not only include time away, but acknowledge the consequences of that leave for subsequent evaluations. As Morgan et al. observed, the productivity difference observed after childbirth would take mothers roughly five years of work to close. It is no surprise, therefore, that taking a leave has a strong negative effect on the likelihood of promotion to full professor (30). It does not benefit mothers to be allowed time away when they will be expected to compensate for that leave when they return.

Impact levels follow similar curves. Impact declines by engagement and is moderated by partner employment, with women in dual roles with an academic partner having higher levels of impact. This is likely a result of the ability to travel and be more visible in service and other professional engagement. Differences in impact, as measured by MNCS, are found to be non-significant for men and women and were in alignment with the time commitment expected from lead-, dual- and satellite parents. They were also not significant for academic couples, suggesting that, at least in part, couples adopt a strategy of cost-minimization to manage potential penalties. In particular, from the qualitative analysis, travel as a way of maintaining visibility in the discipline and maintain collaborations (33), was found to be one of the most difficult academic responsibilities for parents to conduct, where a combination of structured- and unstructured-parenting strategies were necessary to minimize parenting penalty on scientific impact. Scientific organizations should take care to create equity in mobility programs and networking opportunities, to ensure that certain populations are not disadvantaged.

Personal adaptation, however, places the burden upon the scholars and forces adaptations to meet the structural expectations of the "ideal worker". A stronger and more sustainable policy approach would be for institutions to reimagine scientific work to embrace a more diverse workforce. Working demands that assume disengaged parenting disadvantages women. Equitable evaluation requires that institutions consider how their criteria can be applied fairly across populations. These policies, however, should not anchor on the absence of women in

science—that is, through longer leave programs, clock extensions, or virtual programming that allows distant participation. Rather, scientific organizations must imagine more creative ways in which women can be full participants in science. Diversity in science is not only a matter of justice, but is critical for a robust scientific ecosystem (34).

# Figures and Tables

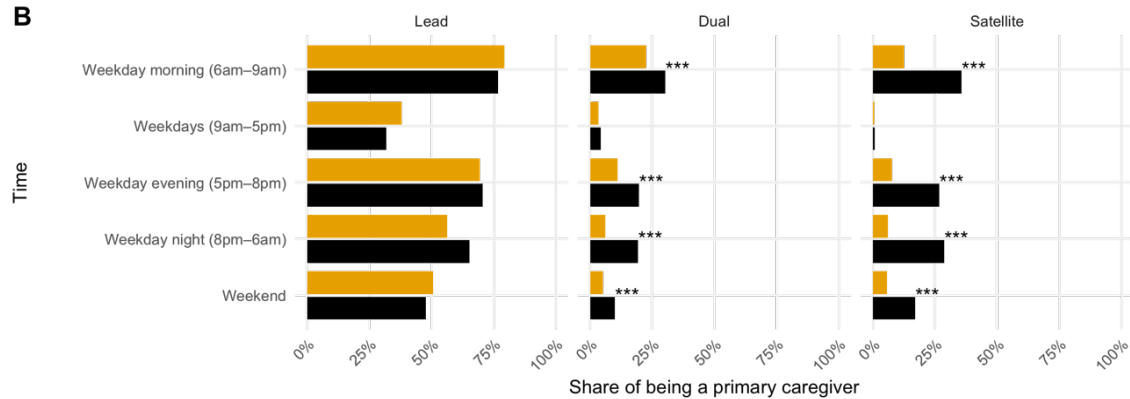

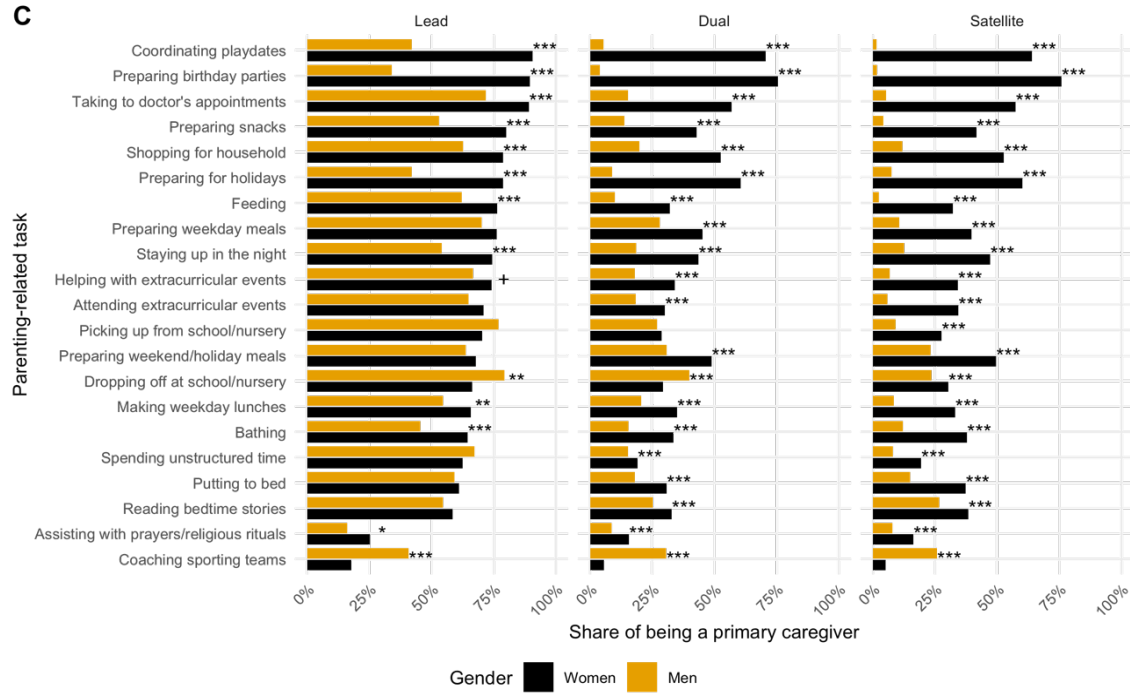

**Figure 1.** (A) Parenting type by gender and household composition. (B) Respondents reporting themselves as the primary caregiver at different times of the day, by gender and parenting type. (C) Respondents reporting being a primary caregiver for the parenting-related activities, by gender and parenting type. Respondents are considered "primary caregiver" if they reported "Mostly me" or "Almost always me" in taking care of these activities. The asterisks denote the FDR-adjusted $p$-values from the two-sample tests of proportion between men and women: + $p<0.1$, * $p<0.05$, ** $p<0.01$, *** $p<0.001$.

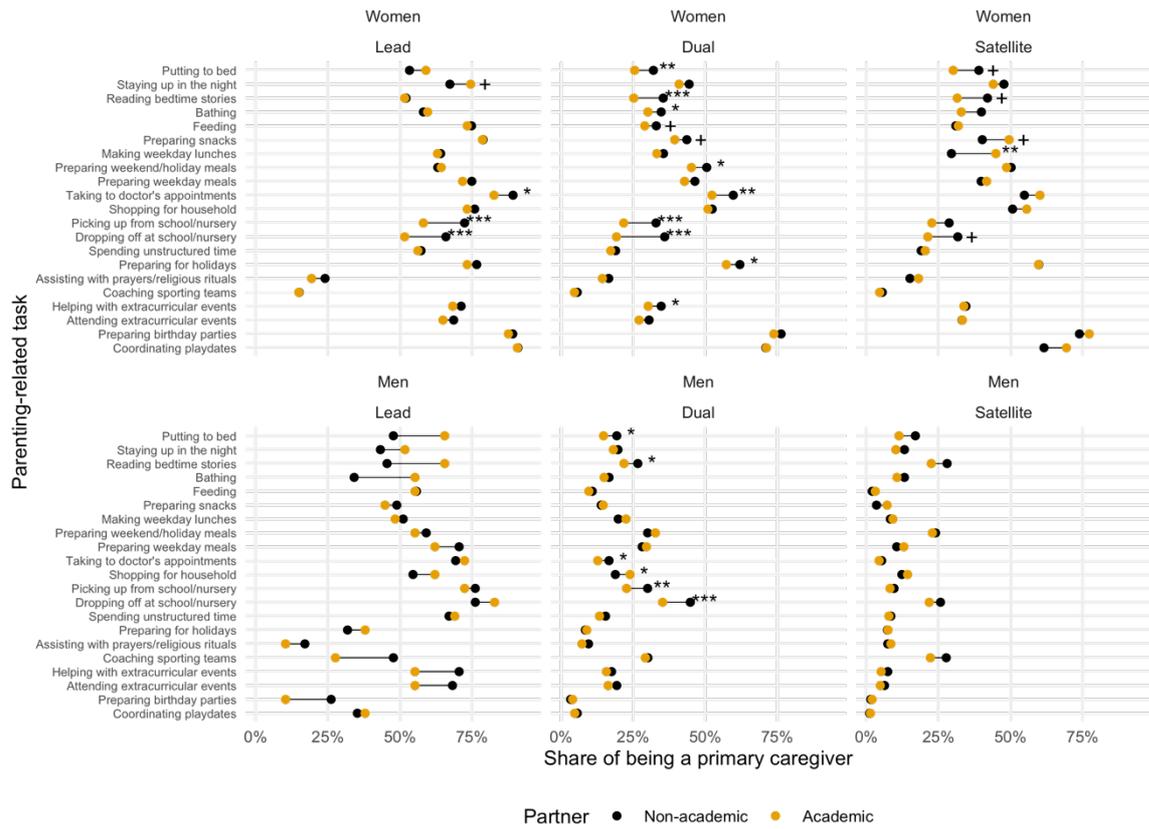

**Figure 2.** Respondents with an academic employment (n=8,046) reporting being a primary caregiver for the parenting-related activities by parenting type, gender, and partner employment status (Academic vs. Non-academic). Respondents are considered "primary caregiver" if they reported "Mostly me" or "Almost always me" in taking care of these activities. The asterisks denote the FDR-adjusted *p*-values from the two-sample tests of proportion between those having an academic partner and their counterparts having a non-academic partner: + *p*<0.1, * *p*<0.05, ** *p*<0.01, *** *p*<0.001.

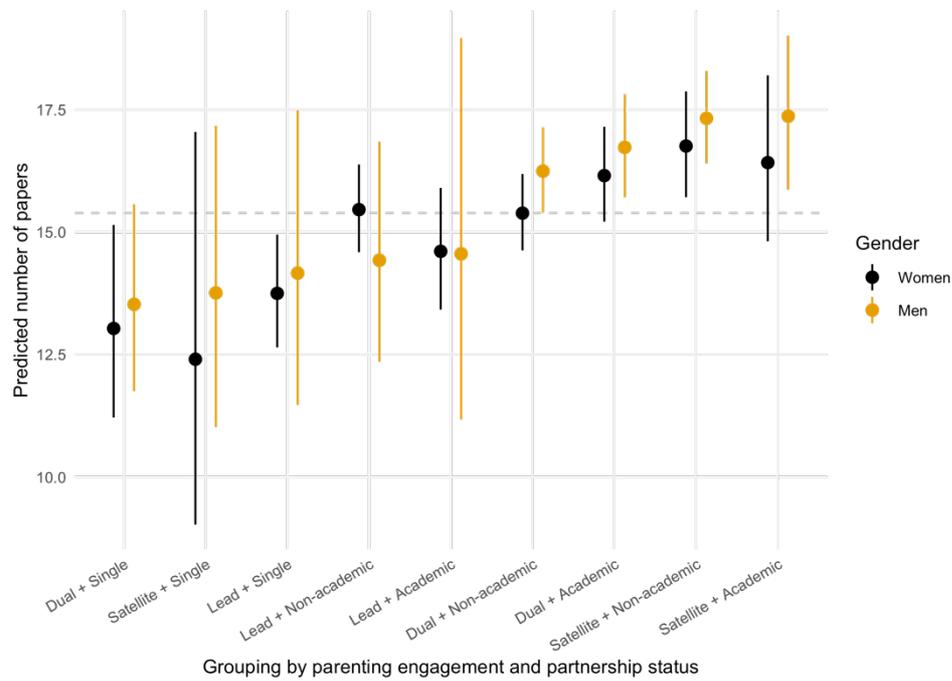

**Figure 3.** Predicted number of papers for men and women, by parenting engagement and partnership status. The dash line refers to the predicted productivity of our reference group: dual mothers with a non-academic partner. Results are averaged over the levels of number of children, academic employment status, doctoral degree, and domain. Intervals are back-transformed from the log scale.

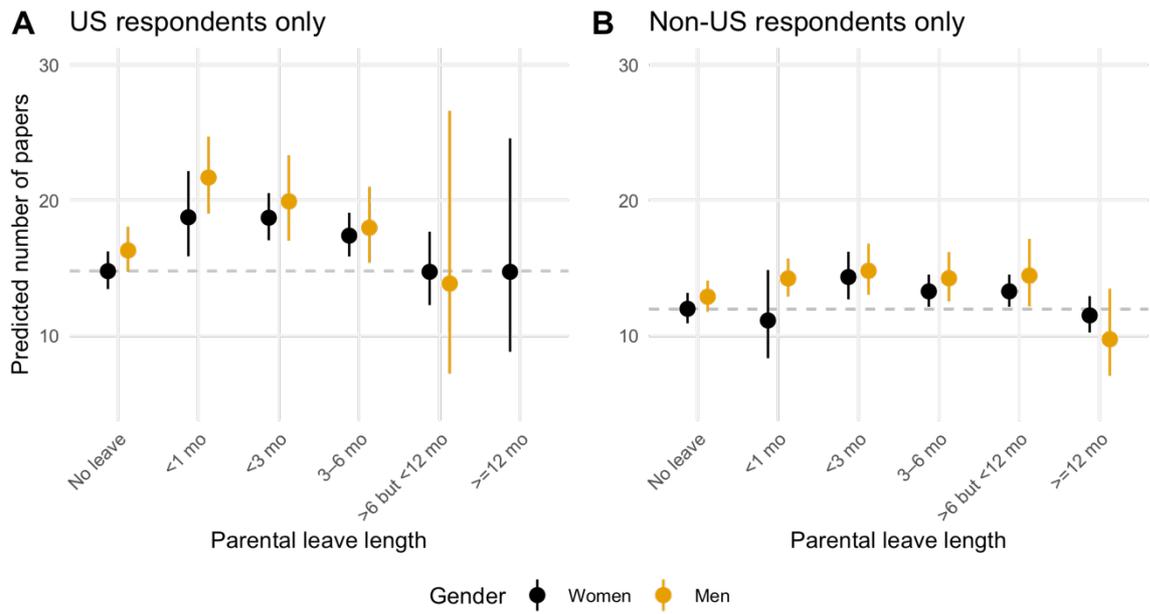

**Figure 4.** Predicted number of papers by parental leave lengths and gender in the (A) US sample and (B) Non-US sample. The dash line refers to the predicted productivity of our reference group: dual mothers with a non-academic partner taking no parental leave. Results are averaged over the levels of parenting type, partnership status, number of children, academic employment status, doctoral degree, and domain. Intervals are back-transformed from the log scale.

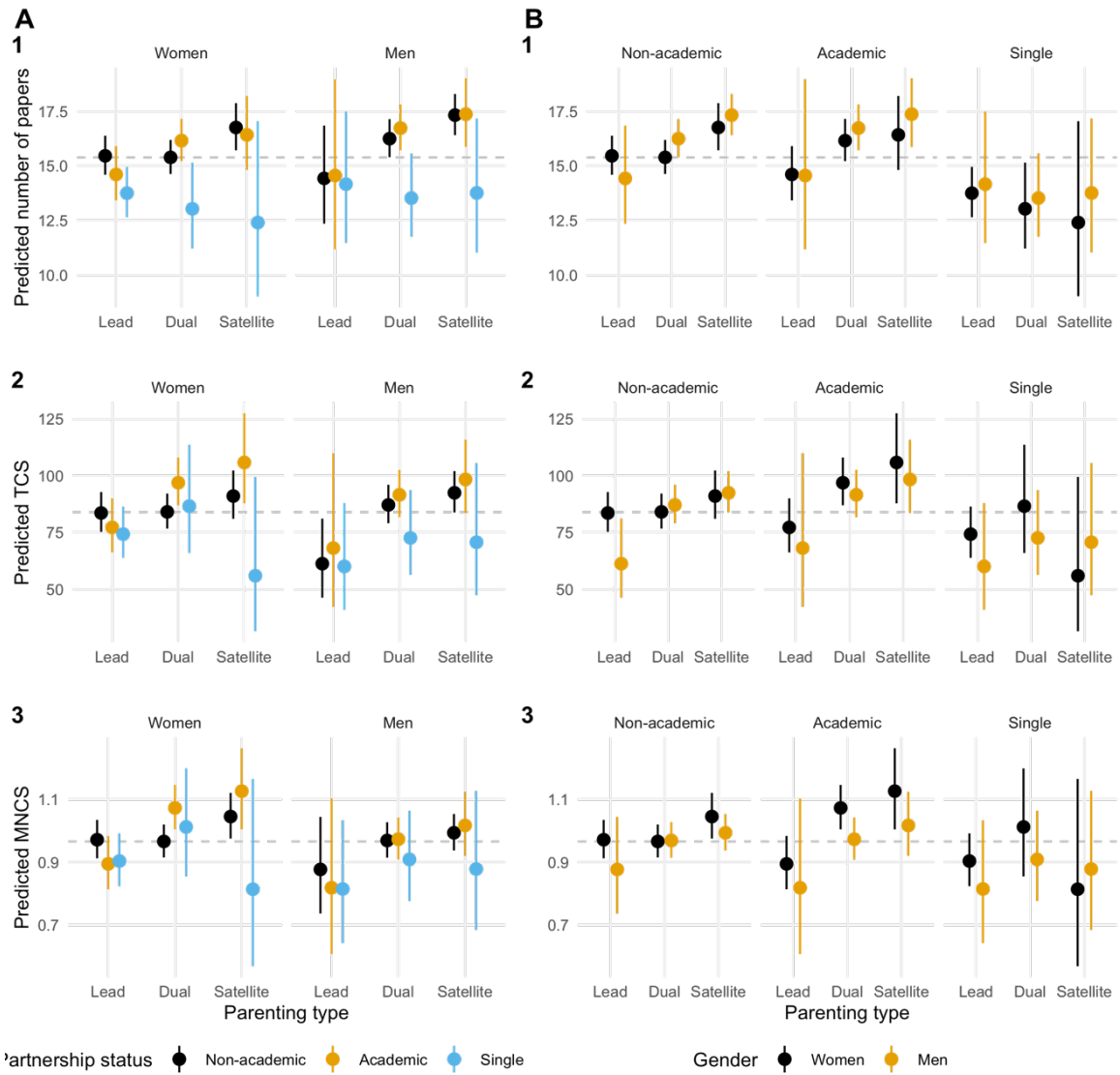

**Figure 5.** Predicted productivity/impact based on parenting type, gender, and partnership status, faceted by (A) gender, (B) partnership status: (1) total number of papers, (2) total number of citations (TCS), (3) mean normalized citation scores (MNCS). The dash line refers to the predicted productivity/impact of our reference group: dual mothers with a non-academic partner. Results are averaged over the levels of number of children, academic employment status, doctoral degree, and domain (except for MNCS). Intervals are back-transformed from the log scale.